\UseRawInputEncoding
\documentclass[pra,12pt,tightenlines,nofootinbib]{revtex4} 
\pdfoutput=1
\usepackage{amsmath}
\usepackage{graphics, graphicx}
\usepackage{epsfig}
\usepackage{ulem}
\usepackage{color}
\usepackage{datetime} 

\begin{document} 


\newcommand{\ttle}[1]{{\it #1}}

\begin{abstract}

``Mind and world ... have evolved together and in consequence are something of a mutual fit.'' Wm. James 

\end{abstract}

\title{What are the Realities?}

\author{James  Hartle}
\affiliation{Department of Physics, University of California, Santa Barbara, California, 93106, USA} {\affiliation{ Santa Fe Institute, \\ 1399 Hyde Park Road,  Santa Fe, New Mexico  87501, USA.}  
\bibliographystyle{unsrt}

\bibliographystyle{unsrt}
\bibliography{references}

\maketitle

\section{Introduction}
\label{intro}
The question in the title of this paper evokes the more frequently asked question ``What is Real?'' Many answers have been give to that question by many deep and distinguished authors in many papers and books e.g \cite{vanFrassen}.  There are even contributions  by the author of this paper \cite{HarQPHL,HarOneReal}.  But, by and large, I have not read the relevant literature on this subject.  Indeed, I am  on record  in  \cite{HarQPHL} as advocating  not using the word ``real'' in scientific discourse as a route to clarity. This paper  provides  another  analysis of  the question ``What is real?'', and provides one answer  to it through the  Realities   of  information gathering and utilizing systems (IGUSes) within  the Universe.

\section{Information Gathering and Utilizing Systems} 
\label{iguses}

The  term IGUS ({\it Information Gathering and Utilizing System})  was coined by the  author and the late Murray Gell-Mann in our work on understanding the application of quantum theory to closed systems like our Universe could  be  \cite{IGUSrefs}. `Observers' played a central role in the Copenhagen formulation of laboratory measurements found in many physics textbooks. But `observers'  and `measurements' could not  be central in a quantum theory of the early Universe where neither existed. So Murray and I  formulated  a generalization of Copenhagen quantum mechanics using the more general idea of an IGUS  to replace human observers in the formulation of a quantum theory of the Universe. This was decoherent or consistories quantum mechanics  e.g. \cite{GH90} and many further  papers available on arXiv.

\vskip .2in

\eject

{\bf IGUSs are approximately localized subsystems of the Universe characterized by the following three properties:
\begin{itemize}
\item{¥} IGUSes acquire information about their environments.
\item{¥} IGUSes use the regularities in the acquired information to create and update a model of their environments and possibly  beyond called its {\it schema}.
\item{¥}IGUSes act on the predictions of this schema, exhibiting behavior, typically acquiring new information in the process.
\end{itemize} }

There are many  familiar examples of IGUSes: As human observers of the Universe  we are IGUSes  \cite{IGUSrefs}. 
 Individual humans are IGUSes, and so are societies of human beings. The humans pursuing science today constitute the human scientific IGUS (HSI) .  Cockroaches and bacteria qualify as IGUSes as would any system that would be called ÔlivingÕ. A self-driving car is an IGUS, so is an airplane on autopilot, a thermostat, etc.,etc. A  robot might be programmed to act as an IGUS  e.g. \cite{HarPN}. There are many other examples.

All these examples of IGUS'es can be seen as emergent features of the Universe's  quasiclassical realm --- the wide range time, place, and scale that the deterministic laws of classical physics hold to a good approximation in our quantum Universe that is characterized fundamentally by indeterminacy and uncertainty \cite{HarQCR} .

The notion of IGUS has  proved useful in all sorts of discussions -- not just those in the author's work on quantum cosmology.   I believe the reason is that the concept of an IGUS has many of the attributes of a human observer but leaves  out those like consciousness that have proved difficult to characterize in physical terms \cite{Gaz}.

Iguses have even found their way into French literature as a poem in Houllebecq's, {\it Possibility of an Island} (trans)  
\cite{PossofIsland}
\vskip.1in 
\centerline{\parbox{5cm}{ We bear witness to,\\
The Apperceptive Center,\\ 
To the Emotional IGUS,\\
Surviving the shipwreck. }}
\vskip  -.1in

\subsection{A Model IGUS} 
\label{model}

\centerline{\includegraphics{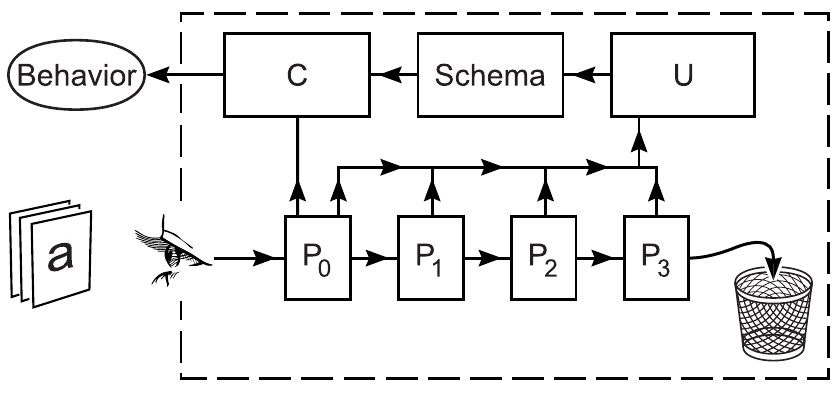} }
 In this section we present a more detailed model of the workings of a  robot IGUS   drawn from \cite{HarPN}.  Its essential features are illustrated in the figure above.  

The information flow in the robot is represented schematically in the  diagram by the arrows. The internal workings of the robot are within the dotted box; its external environment is without. At equal  proper time intervals  the robot captures an image of its external environment. In the example illustrated, this is one of a stack of cards labeled a,b,c, etc. whose top member changes from time to time. The captured image is stored in register P0 which constitutes the robotÕs present. Just before the next capture the image is P3 is erased and images in P0, P1, and P3 are shifted to the right making room for the new image in P0. The registers P1, P2, and P3 therefore constitute the robotÕs memory of the past. The robot uses the images in P0, P1, P2, and P3 in two processes of computation: C (ÒconsciousÓ) and U (ÒunconsciousÓ). The process U uses the data in all registers to make and update a simplified model or schema of the external environment.  That is used by C together with the most recently acquired data in P0 to make predictions about its environment to the future of the data in P0, make decisions, and direct behavior accordingly.




\section{The Realities of IGUSes}
\label{realities} 

In our discussion of Realities  we will restrict to IGUSes composed both individually and collectively of human beings that  have schemas.  It's plausible, even likely, that animals, and even some computers might also have such a notion,  but the author does not want to get bogged down in a discussion of whether an ant ant colony has a schema  although it is certainly an IGUS.  (I don't  know enough about ants.)

Schemata are found  generally in  complex adaptive systems such humans, animals, societies, tribes, etc.   Schemata are essential for the IGUS to predict the future and calculate its behavior. 
But not  every IGUS develops a schema. Many computers gather data but do not construct a schema. 

For those IGUSes that do create a schema it is  possible to define what we shall call 
the {\sl Reality of an IGUS}. {\bf  An IGUS's  Reality consists of those features of its schema that it can rely on in the calculation of productive behavior.}\footnote{I toyed with the idea of calling this {\it  igusic-reality}  but thought better of it.}  
 Since there are many IGUSes  operating in different environments there will generally be many examples of the Realities of an IGUS.  That is another reason why the question `What is Real' is difficult to answer.

\section{Examples of Realities from the Schemata of IGUSes}
\label{realsch}.
\centerline{\bf This section very briefly describes a number of Realities of IGUSes}

\subsection{Physical  Realities}

Examples of physical realities would  include the the every-day realities of tables and chairs, the realities of the arrows of time, the expansion of the universe, the laws of  physics themselves, the big-bang, and our Universe's quasiclassical realm  e.g. \cite{HarQCR}. 

Equipped with confirmed physical theories the HSI can extrapolate its present Reality into the Realities in the future and the past. 

\subsection{Scientific  Realities} The Reality of he Human Scientific IGUS (HSI) includes certain physical realities  such as records of observations and how they change over time, descriptions of how these were obtained --- any apparatus involved  and descriptions of theories that organize and explain all that. The HSI  also generally count as real  the alternatives that are predicted by tested, accepted physical theory.

Was the big-bang real? Most members of the HSI would say ``yes'' because it follows from a confirmed physical theory:  Einstein's general theory of relativity.

Is the anti-deSitter space postulated in the AdS-CFT cprrespondence  real even though we can't go there or make direct measurements of standard classical kinds. 

\subsection{Mathematical Realities}
 Mathematical realities arise from the agreement
by   IGUSes  of mathematicians on axioms which, as G\"odel put it \cite{Godxx},`force themselves on us as being true' A mathematical Reality includes all that can be proved from the assumed axioms.  Different axioms mean different mathematical Realities. 

The author believes that most mathematicians would agree with the above. But many mathematicians act as though there is a reality of mathematical truth somewhere out there and their job is to discover it \cite{MathExp}. When we encounter other intelligent beings we can  predict that
they will have the same arithmetic that we do.  Very probably they will have the  equivalent of the Zermelo-Fraenkel Axioms  of set theory (ZF).  But will they have assumed the axiom of choice (C)?.

\subsection{Historical Realities}
The IGUS of human historians aims at understanding  what ``really'' happened in the past from present records of data found around the world that is  extrapolated to the past with our best theories of evolution in time. 

The lGUS of human cosmological scientists aim at inferring  the  reality of what happened in the physical  past of our Universe from present data of observations on large distance scales. In a quantum Universe there could be many different pasts that could be inferred. \cite{Harpasts}.

\subsection{Realities in Fiction}
As W.H.  Auden  wrote É\cite{auden}  ``the demands one can make of the novelist [are] that he show us the way in which a society works, that he show us an understanding of the human heart, that he create characters whose reality we believe and for whose fate we care, that he describe things and people so that we feel their physical presence, ... ''.
In short the novelist should create a fictional schema  --- a model Reality  of the world that can influence the behavior of some human IGUSes. 

\subsection{Fictional Realities in Science}
One has only to mention the Gibbs ensemble and thought experiments to confirm that there are fictional realities in at least some parts of the schema of the human scientific IGUS.  

\subsection{Faith Based Realities} 
This essay would not be complete without mentioning that that the schema of many human IGUSes contain assumptions that are based on faith. These are therefore parts of the IGUS's reality as we have defined the term. Scientific IGUSes  generally use only the scientific parts of their schema in directing their behavior.

\subsection{The Evolution of Realities} 
\label{evolreal}
The Realiity of  of an IGUS evolves in time. They evolve as the IGUS acquires new information and uses to update its  schema. They evolve as new ideas evolve and as the IGUS moves to new environments, and makes new observations.  Thus the Reality of an IGUS can change in time. 
Realities changing  in time will not be a surprise to human IGUSes. Conscious decisions are often the reason for the evolution. 

Some realities of the Universe seem unlikely to change significantly in time.  For example careful observation along with tested gravitational theory (e.g. the Einstein equation.) suggest that a reality of our Universe is that it will continue to be approximately homogeneous and isotropic on distance scales above several hundred megaparsecs for several billion years into the future.

\section{Reality from Observations}
\label{real-obsd}
The late John Wheeler thought that reality was created by the observations of observers.  For example summing up in his autobiography he says: ``The Universe and all it contains may arise from the myriad of yes-no choices of measurements''   ... it is registration whether by a person a, device, or .. that changes potentiality into actuality'' \cite{Wheelerquotes}.


I discussed his ideas with him many times while visiting him at  his summer home on High Island , Maine. But I could never see a concrete way of realizing them. By and large reality from observations remains in Johnny's phrase ``an for an idea''. 
However  the discussion of Realities in this paper could be said to a fulfillment of his ideas in one way: Realities are in the schemata of IGUSes. An IGUS's schema is constructed, at least in part,  from the results of its measurements.

\section{Emergent Reality}
\label{energ-reak}  

The late Murray Gell-Mann characterized {\it emergence} to mean ``You don't have to do something more to get something more''. \cite{emerg}. The realities defined in this paper are emergent properties of the Universe in that sense. They are found in the schemata of its IGUSes.
A consequence of this discussion is that realities are emergent classical features of our Universe.  That is because IGUSes themelves are emergent features 
of our quantum Universe along with their schemata that define Realities.  A construction or postulate of what's real  in the Universe is not necessary. It's  already there in the schemata of 
its IGUSes.  

Was the big bang real?  Yes, because it is tagged as real in the schema of the human scientific 
Are all the histories in a decoherent set of alternative histories  real? Yes,  because they are found in the schemata of the Human Scientific IGUS working on quantum cosmology, etc.  

\section{Conclusion}  
\label{conclu}
The string of words ``What is Real?'' would seem to define a question that might be answered.. But in fact it is not a question because it does not specify the  ensemble  of possibilities  that  might be real and might not be real. It is not an empirical question but a philosophical and speculative one that might have many answers.  However the question of  what what are the realities specified by the schemata of IGUSes  at any one time  here on Earth or elsewhere in the Universe  is not beyond all hope of  being answered  by physical theory, observation, experiment, history,  records, etc.  

There would  then  be not one just one reality but many different realities that are constructed  from the schemata of many IGUSes. Reality is not as sometimes said ``What's out there independent of human cognition''.  It's more like ``what's in there as a consequence of human cognition and observation''.Therefore we  should not be asking the question `What is Real' as though there is only one thing when there are many Realities.

Perhaps philosophy has answered or will answer the question `What is Real?' in some other way.  But as scientists let us rather try to answer the question `What are the Realities of the IGUSes in our Universe and how do they change in time,?' That is an empirical question which can be answered by standard scientific methods of observation and test. In my opinion, that is a meaningful scientific formulation of the question. `What is Real?


\vspace{.2in}

\noindent{\bf Acknowledgments:}  The work reported in this paper was supported over many years by the US National Science Foundation. The preparation of this   particular paper was supported by  NSF grant PHY-18-8018105. The author has benefited over the years with many discussions  with Mark Srednicki, the late Murray Gell-Mann, and many other colleagues.

\end{document}